\documentclass[a4paper]{spie}  

\usepackage[]{graphicx}

\title{HEXIT-SAT: a mission concept for X-ray grazing incidence telescopes
from 0.5 to 70 keV} 

\author{Fabrizio Fiore\supit{a}, Giuseppe C. Perola\supit{b}, Giovanni Pareschi\supit{c}
Oberto Citterio\supit{c}, Alberto Anselmi\supit{d} and Andrea Comastri\supit{e}
\skiplinehalf
\supit{a}INAF Osservatorio Astronomico di Roma, via Frascati 33, Monteporzio, Italy I00040\\
\supit{b}Universita' Roma Tre, via della Vasca Navale 84, Roma, Italy I00100\\
\supit{c}INAF Osservatorio Astronomico di Brera, Via E. Bianchi 46, Merate (Lc), Italy I23807\\
\supit{d}Alenia Spazio, Strada Antica di Collegno 253, Torino, Italy I10146\\
\supit{e}INAF Osservatorio Astronomico di Bologna, via Ranzani 1, Bologna, Italy I40127
}

\authorinfo{F.F.: E-mail fiore@mporzio.astro.it}

\def\ls{{_<\atop^{\sim}}}
\def\gs{{_>\atop^{\sim}}}
\def\cgs{ ${\rm erg~cm}^{-2}~{\rm s}^{-1}$ }

\begin{document} 
  \maketitle 

\begin{abstract}
While the energy density of the Cosmic X-ray Background (CXB) provides
a statistical estimate of the super massive black hole (SMBH) growth
and mass density in the Universe, the lack, so far, of focusing
instrument in the 20-60 keV (where the CXB energy density peaks),
frustrates our effort to obtain a comprehensive picture of the
SMBH evolutionary properties.  HEXIT-SAT (High Energy X-ray Imaging
Telescope SATellite) is a mission concept capable of exploring the
hard X-ray sky with focusing/imaging instrumentation, to obtain an
unbiased census of accreting SMBH up to the redshifts where galaxy
formation peaks, and on extremely wide luminosity ranges.  This will
represent a leap forward comparable to that achieved in the soft
X-rays by the Einstein Observatory in the late 70'. In addition to
accreting SMBH, and very much like the Einstein Observatory, this
mission would also have the capabilities of investigating almost any
type of the celestial X-ray sources.  HEXIT-SAT is based on high
throughput ($>$400 cm$^2$ @ 30 keV; $>1200$ cm$^2$ @ 1 keV), high quality
(15 arcsec Half Power Diameter) multi-layer optics, coupled with focal
plane detectors with high efficiency in the full 0.5-70keV
range. Building on the BeppoSAX experience, a low-Earth, equatorial
orbit, will assure a low and stable particle background, and thus an
extremely good sensitivity for faint hard X-ray sources. At the flux
limits of 1/10 $\mu$Crab (10-30 keV) and 1/3 $\mu$Crab (20-40 keV)
(reachable in one Msec observation) we should detect $\sim100$ and
$\sim40$ sources in the 15 arcmin FWHM Field of View respectively,
thus resolving $>80\%$ and $\sim65\%$ of the CXB where its energy
density peaks.
\end{abstract}



\section{INTRODUCTION}
\label{sect:intro}  

\begin{figure}
\begin{center}
\begin{tabular}{c}
\includegraphics[height=10cm,angle=-90]{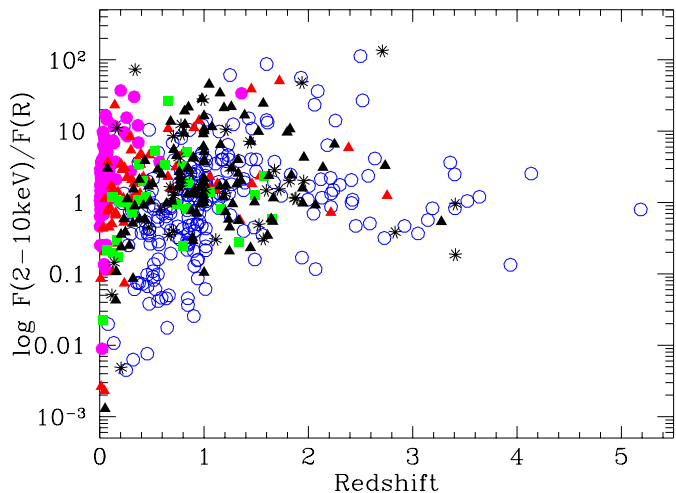}
\end{tabular}
\end{center}
\caption{
\label{fig:xoz}
A projection of
the 2-10 keV discovery space for AGN in the z -- X--ray to
optical flux ratio (X/O) plane. HEAO1= Filled circles; BeppoSAX = filled
triangles; ASCA = filled squares; XMM-Newton = stars;
Chandra = open circles.
} 
\end{figure}

X-ray imaging observations, performed first by Einstein and ROSAT in
the soft X-ray band below $\sim3$ keV and then by ASCA, BeppoSAX,
XMM-Newton and Chandra up to 8-10 keV, have increased by orders of
magnitude the discovery space for both galaxies with an active nucleus
and for thermal plasma sources. As an example, Fig. \ref{fig:xoz}
shows a projection of the 2-10 keV discovery space for AGN in the
redshift versus X--ray to optical flux ratio (X/O) plane.  The use of
collimated detectors on board the UHURU, Ariel-V, and HEAO1 satellites
in the 1970 decade (filled circles in figure) led to the discovery of
$<1000$ X-ray sources in the whole sky, most of which with X/O in the
range 1--10 and very few at z$>0.5$. The first imaging detectors
working above 2 keV on board ASCA and BeppoSAX produced the first
systematic observations of AGN up to z=2-3, and resolved up to 1/4 of
the CXB below 10 keV. The superior image quality of Chandra and the
high throughput of XMM-Newton expanded the discovery space even
further, down to X/O as low as $10^{-3}$, starting to probe the X--ray
emission of star--forming galaxies, and, most important, up to X/O of
several hundred, where many highly obscured, high luminosity type 2
QSO, and many high redshift QSO are found. This allowed us to study
the accretion process on wide ranges of cosmic times, environments and
accretion efficiencies, from very low power AGNs in luminous bulges
(low X/O sources ), to normal AGN (X/O=0.1-10), to highly obscured
QSOs (X/O$\gs10$).  The situation at energies just slightly higher than
8-10 keV contrasts startlingly with this picture. Above 10 keV the
most sensitive observations have been performed so far by a collimated
instruments, the BeppoSAX PDS\cite{Frontera97} (see Fiore, Matt and
Ghisellini contributions at the 2003 ``Restless Universe'' Symposium)
and only a few hundred sources are know in the whole sky, a situation
recalling the pre-Einstein era during the 70'. This limitation
frustrates the possibility to investigate obscured sources (a column
density of $10^{24}$ cm$^{-2}$ produces an optical depth of 1 at 10
keV for solar abundances) and sources dominated by non-thermal
emission. We clearly need to open up a new window in X-ray astronomy
above 10 keV, producing an increase of the discovery space similar to
that obtained with the first soft and medium X-ray imaging missions.
This can be achieved today because the technology necessary to focus
efficiently the hard X--rays is mature. Indeed, several proposal for
an hard X-ray mission have been presented in the last
years\cite{Ferrando03,Pareschi03,Kunieda04,Craig04}.  The purpose of this paper
is to help focusing better the science goals and the needed
technological developments. As a result of this exercise, performed by
a large number of people from the Italian astrophysical and industrial
communities, the HEXIT-SAT mission concept was born. The main
characteristics and architecture of this space project are also
reported in this paper.

\section{Main scientific motivation} 

\begin{figure}
\begin{center}
\begin{tabular}{cc}
\includegraphics[height=7.cm]{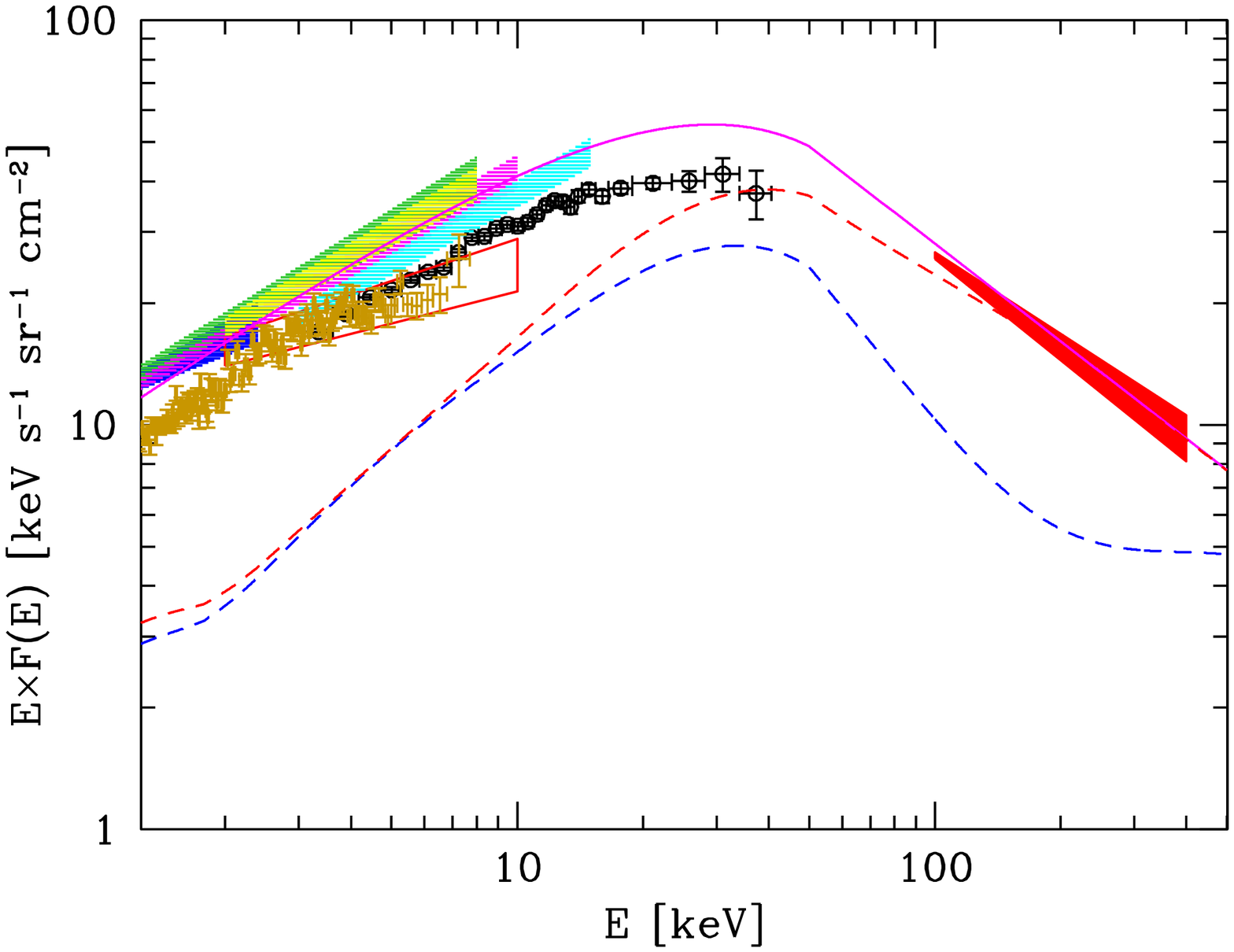}
\includegraphics[height=6.cm]{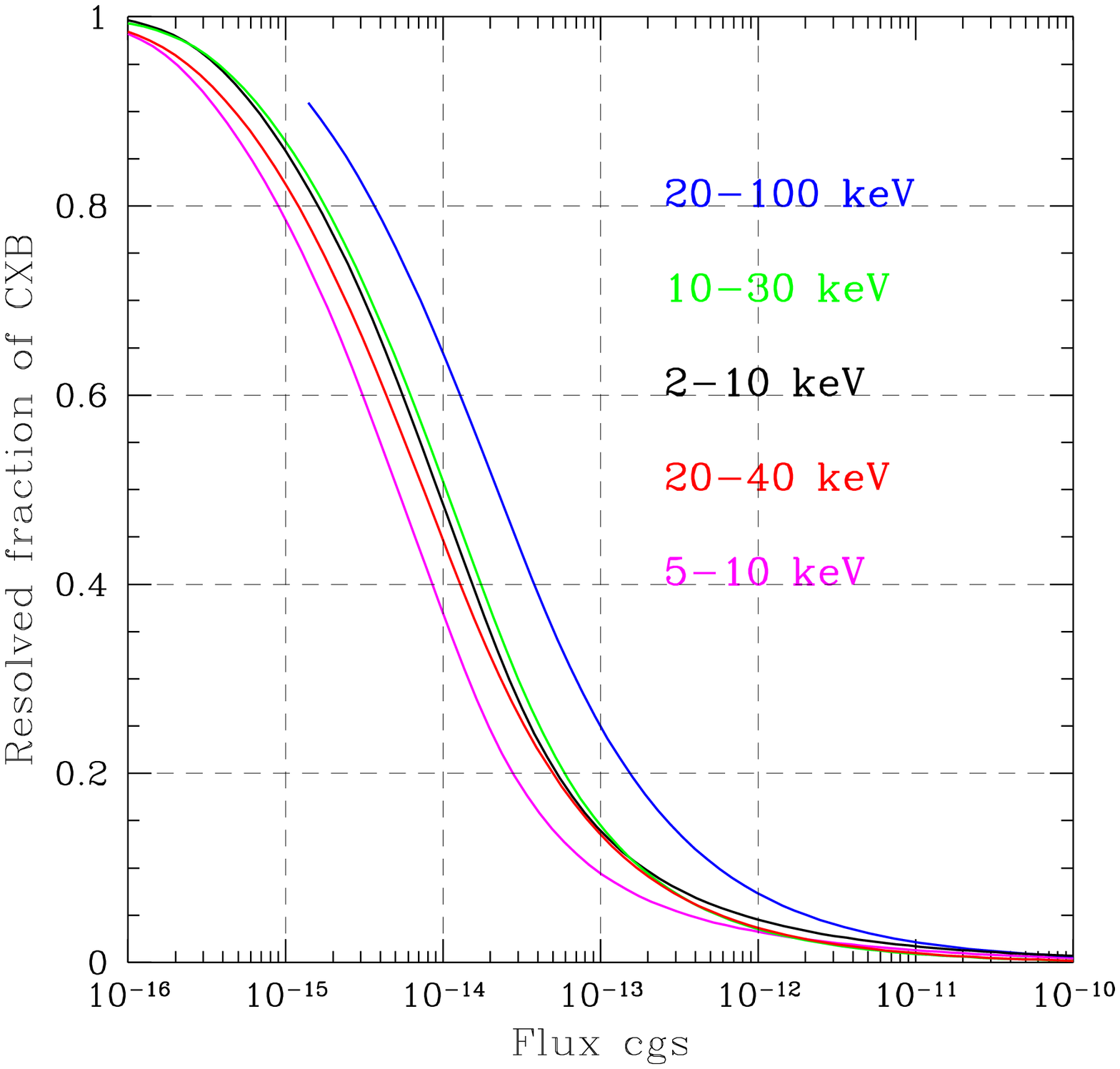}
\end{tabular}
\end{center}
\caption[example] 
{ \label{fig:unres} a), left panel: the residual CXB after subtraction of the
fraction resolved below 10 keV, computed assuming a power law cutoff enegy  of
100 keV (upper dashed curve) and of 400 keV (lower dashed curve). b), right panel:
the fraction of resolved CXB as a function of the flux
computed using the Comastri model. Curves from top to bottom refer to the
bands: 20-100 keV; 10-30 keV; 2-10 keV; 20-40 keV; 5-10 keV.}
\end{figure}

One of the most challenging goals of modern cosmology is to understand
how the structure of the Universe formed and how it evolved with
time. Until ten year ago AGN, shining in about 1/100 of galaxies of
the local Universe, were considered slightly more than a curiosity, in
the framework of galaxy evolution. Indeed, the Cosmic background in
X-rays is about 2 orders of magnitude less intense than that in the
infrared and in the optical.  Two seminal discoveries are now changing
completely this view.  The first is the discovery of SMBH in the
center of most nearby bulge dominated galaxies: their masses are
proportional to bulge properties like mass, luminosity, velocity
dispersion and concentration\cite{Gebhardt00,FM00} The tightness and
steepness of these correlations imply strong interplays and feedbacks
between the nucleus and its host galaxy which in turn imply that if we
want to fully understand galaxy formation and evolution we must
understand the formation and growth of SMBH and AGN evolution.  The
second discovery is that the cosmic history of AGN activity depends on
their luminosity: the number and luminosity density of both low and
high luminosity AGN rises steeply from the local Universe up to
z$\sim1$ but then, while the density of high luminosity AGN stays
constant, that of low luminosity AGN decreases toward higher
redshifts\cite{Hasinger03,Fiore03,Ueda03}. The combined behaviour is
similar to that of the star-formation
rate\cite{Franceschini99,Fiore03}, and indeed, most of the CXB is
thought to be in place at z$<2$\cite{Comastri95,Menci04}.  This
redshift range (z=1--2) can then be considered as the ``golden epoch
of galaxy and AGN activity''.
Conversely, hierarchical clustering $\Lambda$CDM models for the
formation and evolution of galaxies predicts a number of low
luminosity AGN much higher than observed by Chandra and XMM-Newton at z$\gs1$
\cite{Menci04}.  This disagreement may be due to a selection
effect, if highly obscured AGN are common at these redshifts (as in
the nearby Universe) but are missed (or their luminosity is badly
under-estimated) in Chandra and XMM-Newton surveys.  Indeed, only 40--50\% of
the hard (7-10 keV) CXB has been resolved\cite{Worsley04}
leaving room for a population of highly obscured AGN at z$\simeq 1-2$
where the bulk of the CXB is produced.  The spectrum of the
``unresolved'' background can be estimated by subtracting from the
total CXB (normalized to the De Luca \& Molendi 2004 XMM-Newton
determination\cite{DeLuca04}) the contribution of the sources detected by Chandra and
XMM-Newton below 8-10 keV. This is given in Fig. \ref{fig:unres}a) adapted from
Comastri 2004\cite{Comastri04}.  Not surprisingly, this spectrum is reminiscent of that
of extremely obscured sources.  In conclusion, the light-up and
evolution of obscured accreting SMBH is today still largely unknown.
It is clear that for an unbiased census of the AGN population making
the bulk of the CXB and in particular of highly obscured sources
($N_H\gs10^{24}$ cm$^{-2}$, sensitive observations extending up to
about 70 keV are needed.

\subsection{Additional scientific motivations}

Besides the main goal described in the previous section, a mission
with good sensitivity and imaging quality up to 60-70 keV would have
the capability to investigate almost any type of the celestial X-ray
sources, and, in particular it would be crucial for the study of hard
X--ray non-thermal emission from a wide variety of sources.
Acceleration mechanisms in AGN, cluster of galaxies and stellar
sources (both single stars, binaries and Supernovae Remnants) are
today rather poorly known and studied. In the following we give two
examples of forefront science that could highly benefit from such a
mission.

Thanks to the PDS sensitivity BeppoSAX was able for the first time to
observe a few blazars in which the peak of the synchrotron emission
reached around 100 keV (while the peak of the Inverse Compton (IC)
component is in the GeV or even TeV range, the so called High Energy
Peacked sources, HEP). These nearby low luminosity sources are
probably the tip of the iceberg of a much larger population. The about
2 orders of magnitude increase in sensitivity achievable with respect
to the PDS will make feasible both the discovery and the study of a
statistical sample of HEP, and, possibly, the discovery of even more
extreme sources, with the syncrotron peak at even higher energies.

The presence in cluster of galaxies of a diffuse component of
relativistic particles is witnessed by the observation, in a several
of them, of synchrotron diffuse radio emission (so called radio-halo
sources). The inevitable IC interaction of the relativistic electrons
with the photons of the cosmic microwave background should provide
emission in the X-rays, whose intensity would dominate the thermal
emission at E$>10$ keV . Its detection would constrain at the same
time the number of electrons and the strength of the magnetic field
(not possible using the synchrotron emission alone). A positive
result, but still controversial (a few $\sigma$), was obtained by the
BeppoSAX PDS in the Coma cluster and in Abell 2256\cite{FF00,FF04}
but also see\cite{RM04}. The
much increased sensitivity available with focusing optics with respect
to the PDS and, most important, their high quality imaging
capabilities, will allow a detailed mapping of the IC component, and
therefore of the intra-cluster magnetic field and relativistic
particles.

\subsection{Requirements}

Fig. \ref{fig:unres}a) shows that the highly obscured sources
contributing to the ``unresolved'' fraction of the CXB are probably
still a minority population even in the 10-30 keV band, contributing
to $40-50\%$ of the total CXB in that band. In the 20-40 keV band the
contribution of the highly obscured AGN undetected below 10 keV should
be 50-70\% of the total. This means that if we want to discover
sizeble samples of this elusive population and make a fundamental step
forward with respect to the Chandra and XMM-Newton results, we need to
resolve in sources the majority of the 10-30 keV CXB, say 80\%,
similar to what Chandra and XMM-Newton did in the 2-10 keV band, and
at least 50\% of the 20-40 keV CXB.  From Fig. \ref{fig:unres}b),
showing the fraction of the CXB resolved in sources as a function of
the flux in different energy bands calculated using the Comastri et
al.  AGN synthesis model of the CXB\cite{Comastri95}, we see that the
above goals can be reached if we can push the observations in the
10-30 keV band to $2\times 10^{-15}$ \cgs (corresponding to $\sim0.13
\mu$Crab) and to $7\times 10^{-15}$ \cgs (corresponding to 3/4 of
$\mu$Crab) in the 20-40 keV band.

\subsubsection{Image quality and confusion limit}

The goal flux limits in the previous section have been computed using
the Comastri et al. prediction in the 10-30 keV and 20-40 keV bands.
We verified that these prediction agree well with those obtained by
Menci et al. (2004) using a completely different theoretical approach,
and with those of Ueda et al. (2004) which extrapolate at harder
energies the best fit to the observed 2-10 keV AGN luminosity
function. Using all these predictions we evaluate in
$\sim2300$deg$^{-2}$ and $\sim350$deg$^{-2}$ the source density at the
flux limits of 1/10 and 3/4 $\mu$Crab.  Using the standard criterion
for source confusion we calculate the probability P to have two
sources closer that the beam size as a function of the beam size
$\theta_{FWHM}$, assuming the higher source density. We find that
Point Spread Functions (PSFs) with Half Power Diameter (HPD) of 15, 30
and 50 arcsec (and with HPD=2$\theta_{FWHM}$, which is typical of
realistic grazing incidence mirrors, because of the power lost in the
broad scattering wings), give probabilities of 10\%, 30\% and 50\%
respectively. These should actually be considered as lower limits
because the classical analytical formula of course does not account
for the efficiency of the detection algorithm, the PSF sampling, and
source clustering.  This means that a mirror PSF with HPD$\ls15-20$ is
necessary to avoid strong source confusion (P$<10-20\%$) at the flux
limit of 0.1 $\mu$Crab.

\section{Outlining a mission concept: the payload}

\subsection{Telescope design tradeoffs}

A flux of 0.1 $\mu$Crab gives 0.057 counts/Msec/cm$^2$ in the 10-30 keV
band and therefore an effective area of $\gs350$ cm$^2$ @30 keV is
needed to collect 20 counts in 1Msec, which would give a 3$\sigma$
detection assuming a similar background count rate. 
the mirror effective area is given by:

$$ A_{eff}\sim FL^2 \times \theta^2 \times R^2 $$

where FL is the telescope focal length, theta is the grazing angle and
R is the mirror reflectivity. Therefore at least two strategies can be
envisaged to obtain large collecting areas at high energy: a) very
long focal lengths with usual single-layer reflecting coatings; b)
high mirror reflectivity, which can be obtained using multilayer
coatings on mirror shells with grazing incidence close to the critical
angles that would be achieved for total reflection in the 10 20 keV
energy band, e.g. 0.2-0.3 deg. For our exercise we followed the second
strategy. An important advantage in keeping not too high the focal
length (less than 10 m) is a small focal spot (due to a more reduced
plate-scale) with a consequent decrease of the intrinsic background
counts. In addition , another fundamental advantage of this strategy
is that sufficiently large reflection angles, as permitted by
multilayer mirrors, allow one to obtain a still relatively large Field
of View (FOV), of the order of 15 arcmin diameter FWHM. This is
important because the source densities at the goal flux limits in the
10-30 keV and 20-40 keV bands are 120 and 12 sources per 15arcmin
diameter FOV respectively.

\subsection{The X-ray Mirror Modules}

\begin{figure}[ht]
\begin{center}
\begin{tabular}{cc}
\includegraphics[height=7cm]{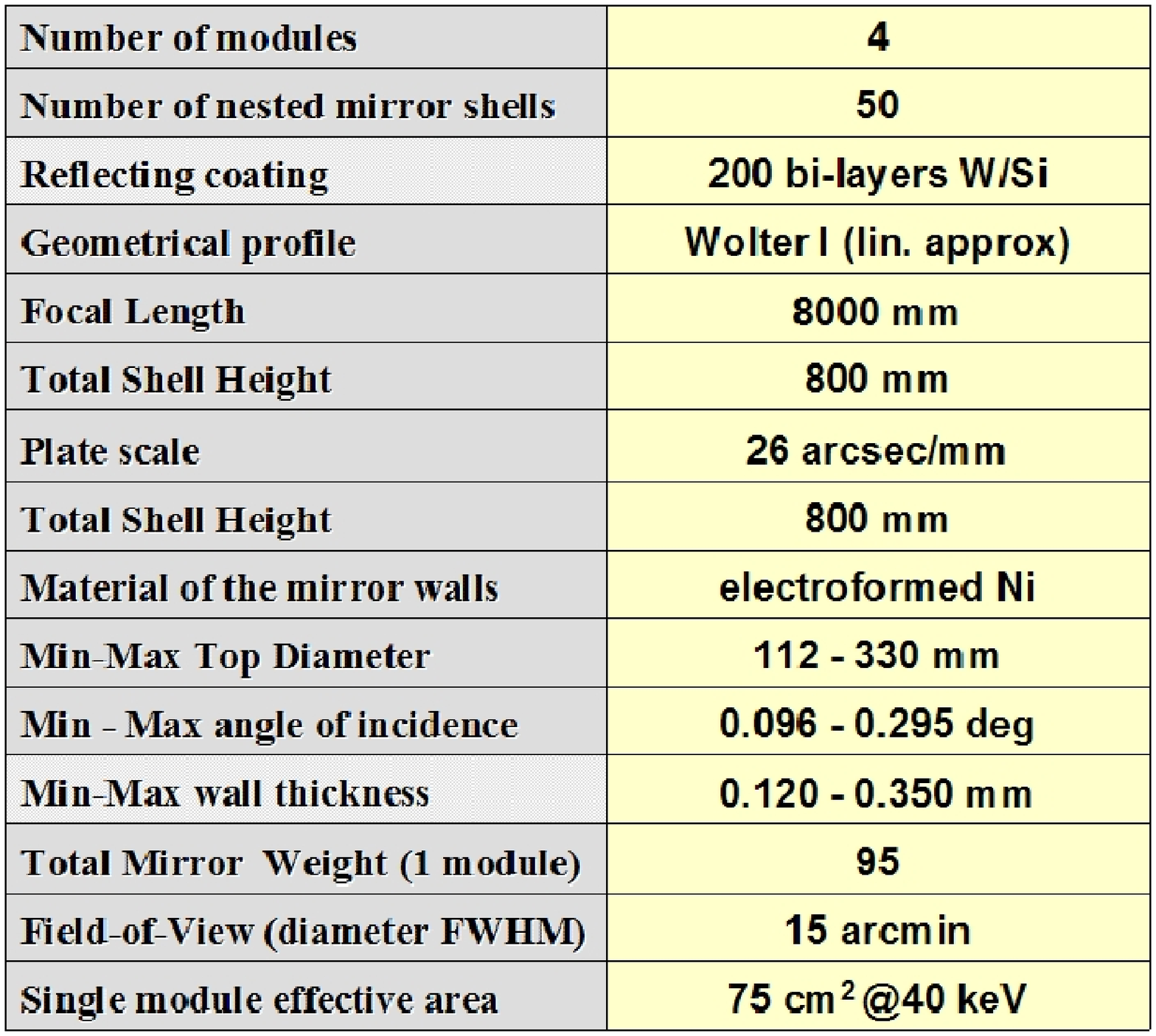}
\includegraphics[height=7.3cm]{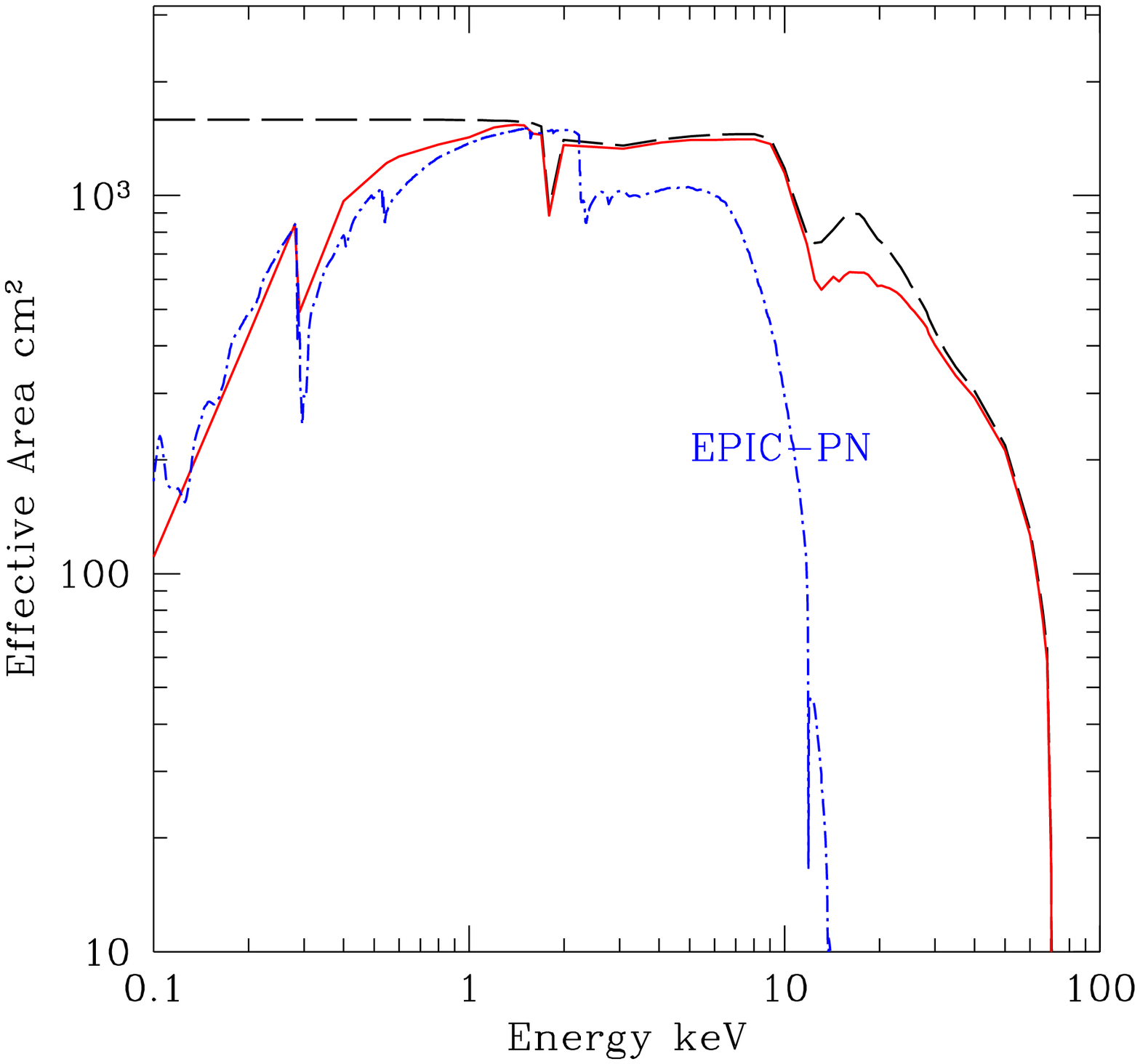}
\end{tabular}
\end{center}
\caption[example] 
{ \label{fig:table} a), left panel: Table with the main characteristics of the
HEXIT-SAT mirror modules. b), right panel: total effective area of 4 modules (long dashed line),
effective area corrected for detector efficiency (solid line) compared with the EPIC-PN effective
area (dot-dashed line).}
\end{figure}

The main characteristics of the mirror modules designed 
for HEXIT-SAT are given in Fig. \ref{fig:table}a). The needed effective area
is obtained using 4 mirror modules, each with 8m focal lenght.
The sequence design for the thicknesses of bi-layers along the
multilayer stacks will continuously change following a power-law, in
agreement with the method adopted by\cite{Joensen95}. The
power-law parameters have been separately optimized for each mirror
shell by a global mathematical approach (the so called "Iterated
Simplex Algorithm" procedure), assuming as a figure of merit the
enhancement of the effective area in an energy band as large as
possible.  The total on-axis effective area for 4 multilayer modules,
not-corrected for the detector efficiencies, is reported in Fig.
\ref{fig:table}b). It should be noted that the cut-off of effective area
around 70 keV is determined by the K absorption edge of W layers.

The fabrication technique for the hard X-ray mirrors is based on
direct replication by Ni electroforming of the multi-layer film
previously deposited onto a mandrel.  This approach is an upgrade,
after appropriate modifications, of the production process (developed
in Italy) successfully used for the Au-coated high throughput optics
with high quality imaging properties of the BeppoSAX\cite{Citterio88},
JET-X/SWIFT and XMM-Newton soft X-ray experiments.  The study to
produce a complete prototype has been financed by the Italian Space
Agency (ASI) and is on going (Pareschi et al. this
conference\cite{Pareschi04}), with the goal of demonstrating the
readiness and the quality of the technology In particular, the
manufacturing method adopted for the production of multi-layer optics
foresees the following steps: a) a superpolished Aluminum mandrel with
an external layer ($\sim$100 $\mu$m thick) of electroless Nickel (Kanigen) is
fabricated. The shape of the mandrel is the negative profile of a
linear-approximation Wolter I mirror to be
realized.  b) on the mandrel surface is then deposited a multi-layer
film by means of a proper deposition technique, able to alternate the
deposition of two different materials, in order to allow the growth of
both materials (spacer and absorber) of the multi-layer stack. The
mandrel is mounted with its axis of symmetry perpendicular to the
deposition beam of evaporated material and it is rotated during the
deposition process. During the film growth the layer thickness is
measured by means of a cooled Quartz microbalance, while the
uniformity of the sputtering beam is obtained using an equalization
mask; c) the mandrel is then put into an electrolytic bath, where a
layer of Nickel is deposited on the multi-layer stack. Afterward the
multi-layer mirror is separated from the mandrel by cooling it,
exploiting the fact that the CTE of the Al (i.e., the bulk material of
the mandrel) is about twice larger than the electroformed Ni walls.
It should be noted that this approach has been already successfully
proved for flats and small-dimension mandrels by using the Ion Beam
Sputtering as a deposition method\cite{Pareschi00}.
The process is now being up-graded with the use of the Ion
assisted E-beam Deposition instead of Ion sputtering, which more
easily permits to extend the use of the technique to large-size
mandrels.

\subsection{Focal plane detectors}

\begin{figure}[ht]
\begin{center}
\begin{tabular}{c}
\includegraphics[height=5cm]{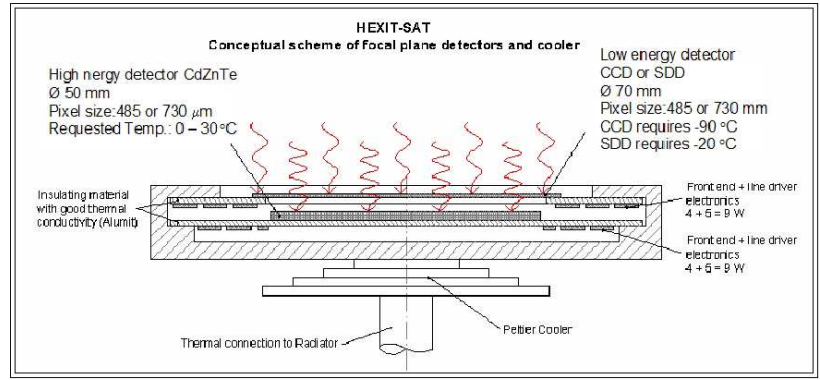}
\end{tabular}
\end{center}
\caption[example] 
{ \label{fig:det} Conceptual scheme of focal plane detectors and cooler}
\end{figure}

The goal to be achieved is to have a detector system that matches well 
the optics performances over the broad  0.5 - 70 keV energy range.
We need a small area
imaging system that has both good spatial and energy resolution. The
angular resolution of the X-ray telescope will be $\sim15$arcsec HPD, 
which requires, in the focal plane, a spatial resolution no larger than
0.2-0.3 mm, to sample the PSF at least with two pixels.  The
detector must also have good efficiency not only in the so called
"hard" X-ray band ($\gs$10 keV) but also it is considered of extreme
importance having an optimal response also in the "classical" X-ray
band (0.5 - 10 keV), where the HEXIT-SAT telescopes present the largest
effective area. We have therefore envisaged two possible solutions:

\begin{itemize}

\item[i)] the first one is based on the use of solid state detectors
covering the 10-70 keV band, placed in series after a CCD or a SDD
detector, which in turn can detect photons between 0.5 and 12
keV. This is the baseline solution adopted in this document for the
sensitivity graphs and textual references \ref{fig:det}.  The hard
X-ray detector choice will be a CdZnTe hybrid solid state detector
(the possibility of using CdTe crystals is an alternative solution).
CdZnTe detectors combine the room temperature operation with a good
spectroscopic performance, theoretically approaching that of Ge and Si
detectors (that, however, would require complex cryogenic
systems). This clearly implies a saving in the overall instrumentation
complexity, cost and weight. Due to the relatively high atomic numbers
(49) and to the high density (6.1 g/cm$^3$), the quantum efficiency is
also very good ($\sim99\%$ for a 1 mm thick crystal @ 55 keV). Thanks
to their low capacitance CdZnTe detectors allow to reach high energy
resolution when an accurate coupling to custom front-end electronics
is used.  The required pixel size of 0.2-0.3mm implies further
developments of the current CdZnTe pixelated detectors.
The realization of a prototype detector is planned in the framework
of the mentioned Pareschi et al. study (also see Del Sordo et al. 2004
\cite{DelSordo04})

\item[ii)] the other approach is instead based upon a SDD
optically connected to an array of microscintillator crystals\cite{Labanti00}.
\end{itemize}

CCD detectors are readily available but require temperatures as low as
-90C. Silicon drift chambers have very good energy resolution at
higher temperature (-20C), but require some development.

\begin{figure*}
\begin{center}
\begin{tabular}{c}
\includegraphics[height=12cm,angle=-90]{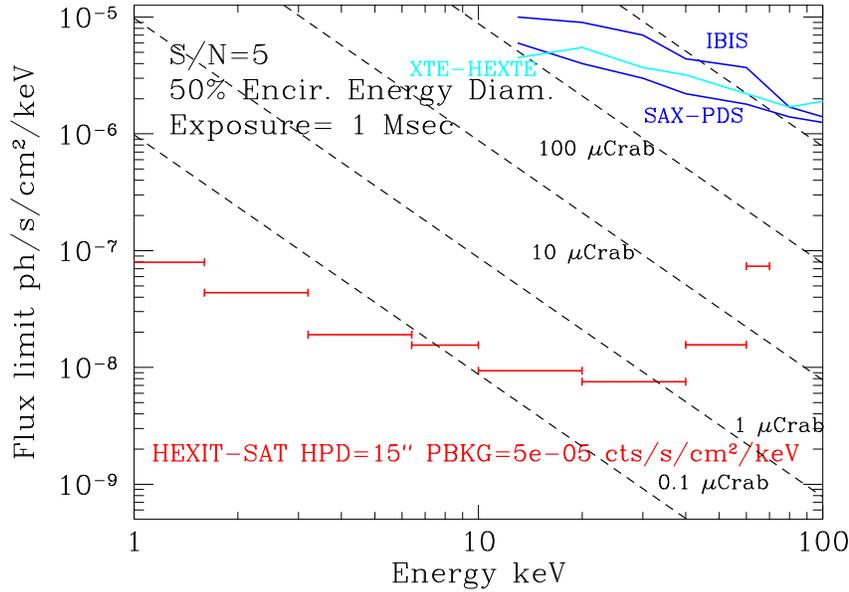}
\end{tabular}
\end{center}
\caption[example] 
{ \label{fig:flimene} S/N=5 flux limit reachable in 1Msec in the 1-70 keV band}
\end{figure*} 

\subsection{Expected performances}

\begin{figure}[ht]
\begin{center}
\begin{tabular}{cc}
\includegraphics[height=7cm]{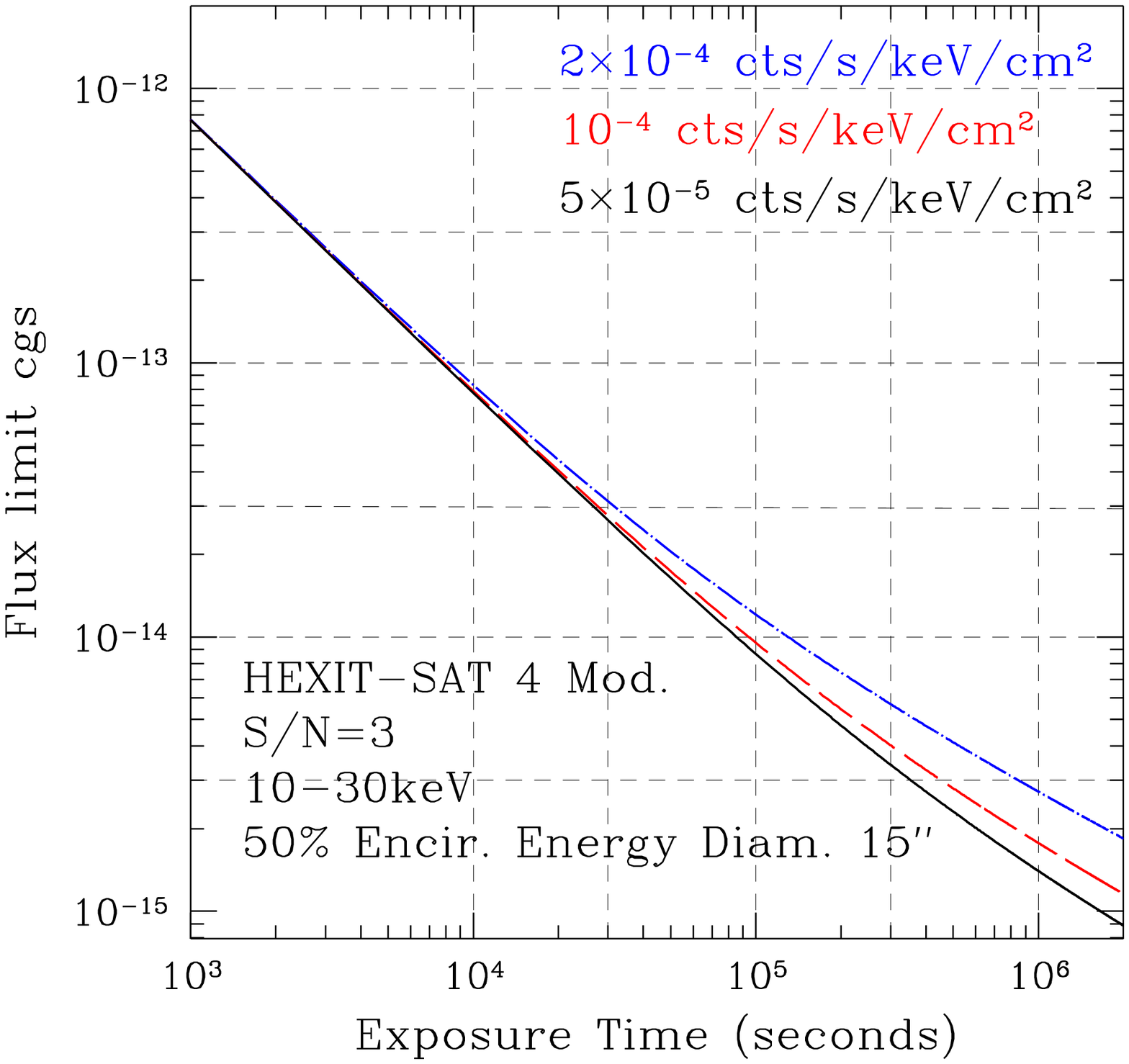}
\includegraphics[height=7cm]{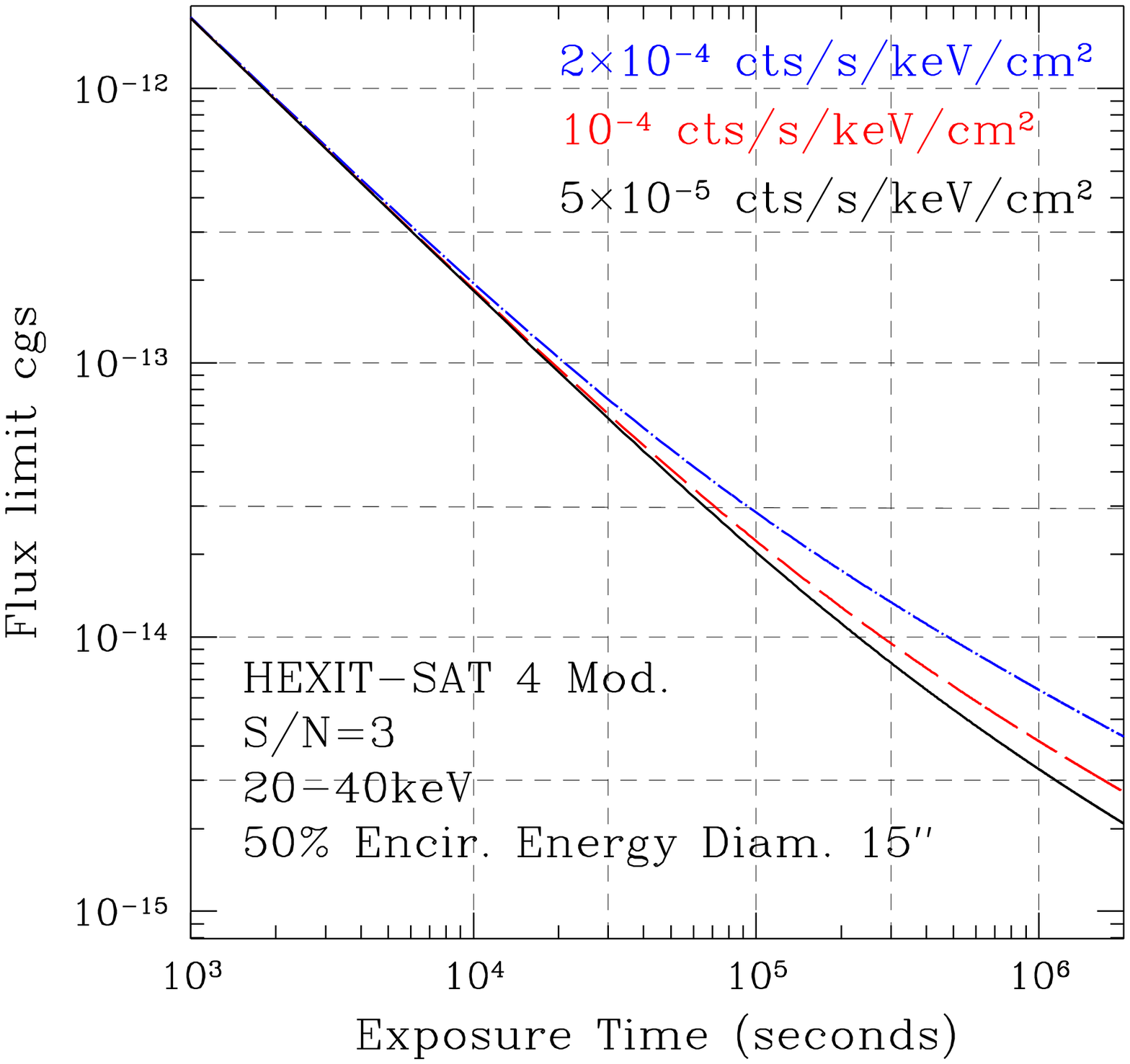}
\end{tabular}
\end{center}
\caption[example] 
{ \label{fig:flimexpo} S/N=3 flux limits in the 10-30 keV (left panel) and 20-40 keV 
(right panel) as a function of the exposure time for three particle induced background
values}
\end{figure}

The design briefly illustrated in the previous section implies that
the deepest HEXIT-SAT exposures will be background dominated. The
ultimate performances of this experiment depend therefore on the
satellite environment and on our ability to reject the different
background components: a) particle background; b) X-ray photons
produced by the interaction of the particle with the spacecraft; the
CXB.  The BeppoSAX PDS proved that one of the best environment is a
low Earth orbit with low inclination, and therefore we will use this
baseline orbit in the following calculations. The background seen in
the BeppoSAX PDS NaI(Tl) detector, after active rejection in the
CsI(Na) detector and in the lateral and top shields, is of
$\sim5\times10^{-5}$ cts/s/cm$^{-2}$ per each mm of detector
thickness.  We assume this value as the reference particle induced
background (direct particle + X-ray induced photons).  Using the
effective area in Fig. \ref{fig:table}b), which also includes detector
efficiency, and this internal background plus the contribution of
the CXB, we computed the flux limits as a function of the energy in
Fig. \ref{fig:flimene}.  Our mission design produces an
improvement in sensitivity of a factor of $\sim$300 with respect to
the BeppoSAX PDS at 30 keV.  Fig. \ref{fig:flimexpo} shows the flux
limit (for a signal to noise of 3) as a function of the exposure time
in the 10-30 keV and 20-40 keV bands. For the baseline
background in 1 Msec we can reach a 10-30 keV flux of
$1.4\times10^{-15}$ \cgs (0.09 $\mu$Crab) and a 20-40 keV flux of
$3.3\times10^{-15}$ \cgs (0.35 $\mu$Crab).  At this flux limits there
should be $\sim120$ and $\sim40$ sources per FOV in the 2 bands
respectively, which correspond to resolving $\gs80\%$ and $\sim65\%$
of the 10-30 keV and 20-40 keV CXB, similar to the goals indicated in
Sect. 2.2.

\begin{figure}[ht]
\begin{center}
\begin{tabular}{cc}
\includegraphics[height=7cm]{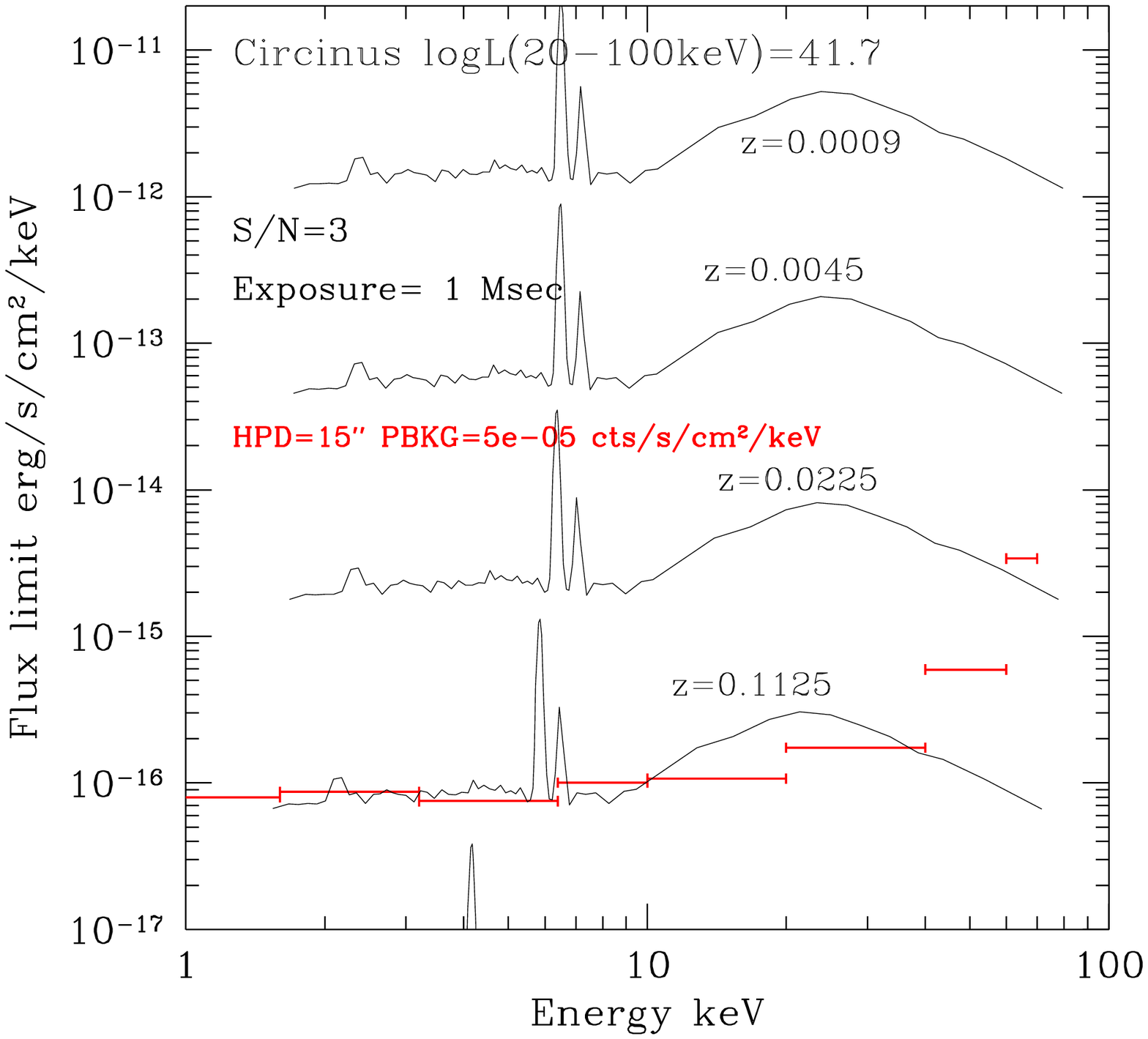}
\includegraphics[height=7cm]{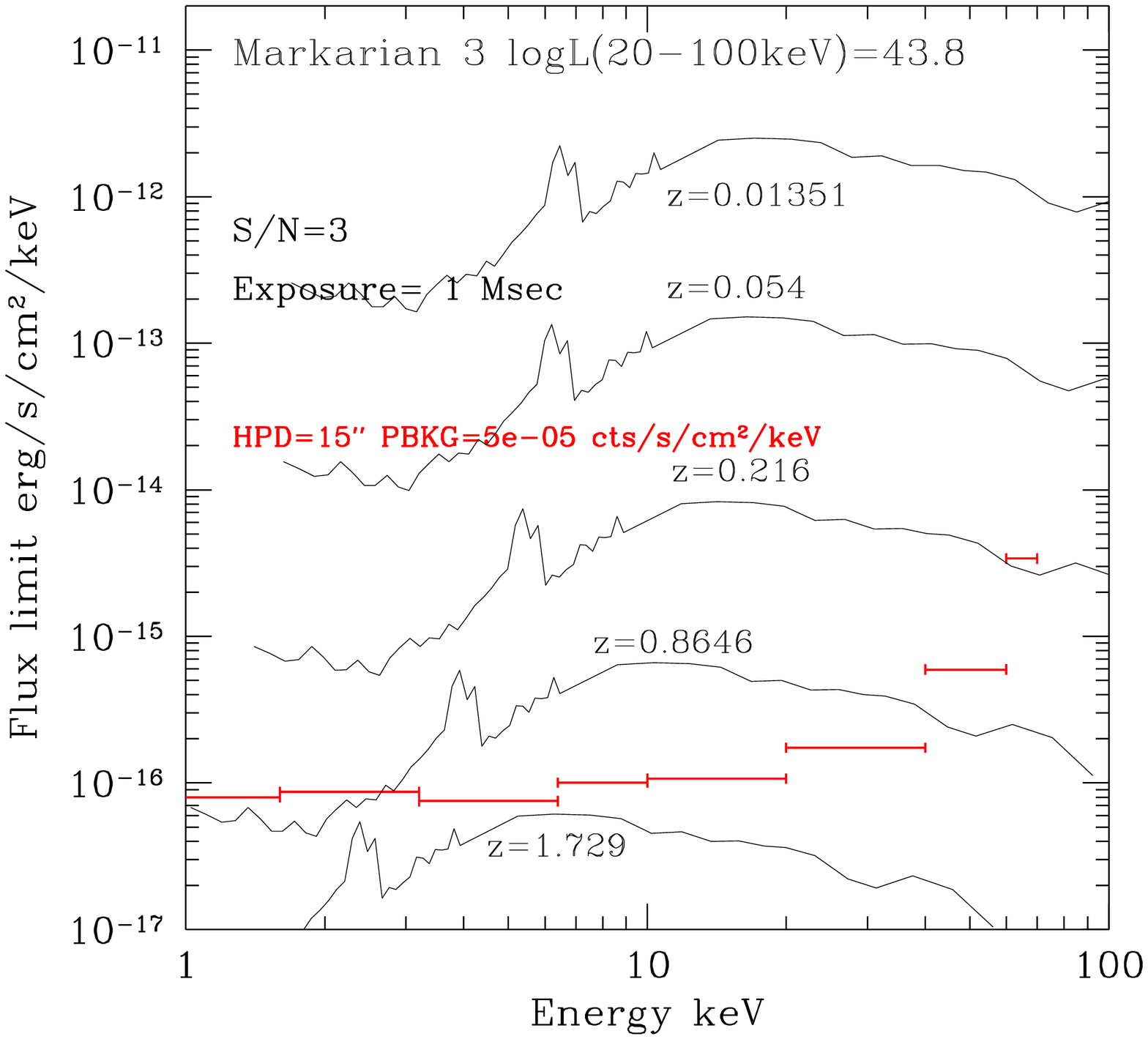}
\end{tabular}
\end{center}
\caption[example] 
{ \label{fig:mark3} HEXIT-SAT flux limits compared to the
redshift spectra of the Seyfert 2 galaxies Circinus Galaxy and
Markarian 3. The spectra of these AGN have been measured accurately in the
full 1-100 keV band by the BeppoSAX MECS and PDS instruments.}
\end{figure}

We tried to understand up to which redshift HEXIT-SAT is able to
detect AGN similar to the highly obscured Seyfert galaxies common in
the local Universe. Two examples are given in Fig. \ref{fig:mark3}.
We have redshifted the spectra of the low luminosity Seyfert 2 galaxy
Circinus galaxy (distance of 4 Mpc, logL$_{2-10keV}=41.7,~
N_H=2\times10^{-24}$) and of the more luminous Seyfert 2 galaxy
Markarian 3 (z=0.0135, logL$_{2-10keV}=43.8,~N_H=2\times10^{-24}$) and
compared with the 1 Msec flux limit of HEXIT-SAT. Objects
similar to Markarian 3 are detectable up to z$\sim1$, i.e. at the peak
of the AGN number and luminosity density, while the low luminosity
Circinus galaxy would be detected up to z$\sim0.1$.  We conclude that
the proposed mission would produce a large increase of the discovery
space for obscured AGN.

\section{Outlining a mission concept: critical issues on spacecraft design}

\begin{figure}[ht]
\begin{center}
\begin{tabular}{cc}
\includegraphics[height=7cm]{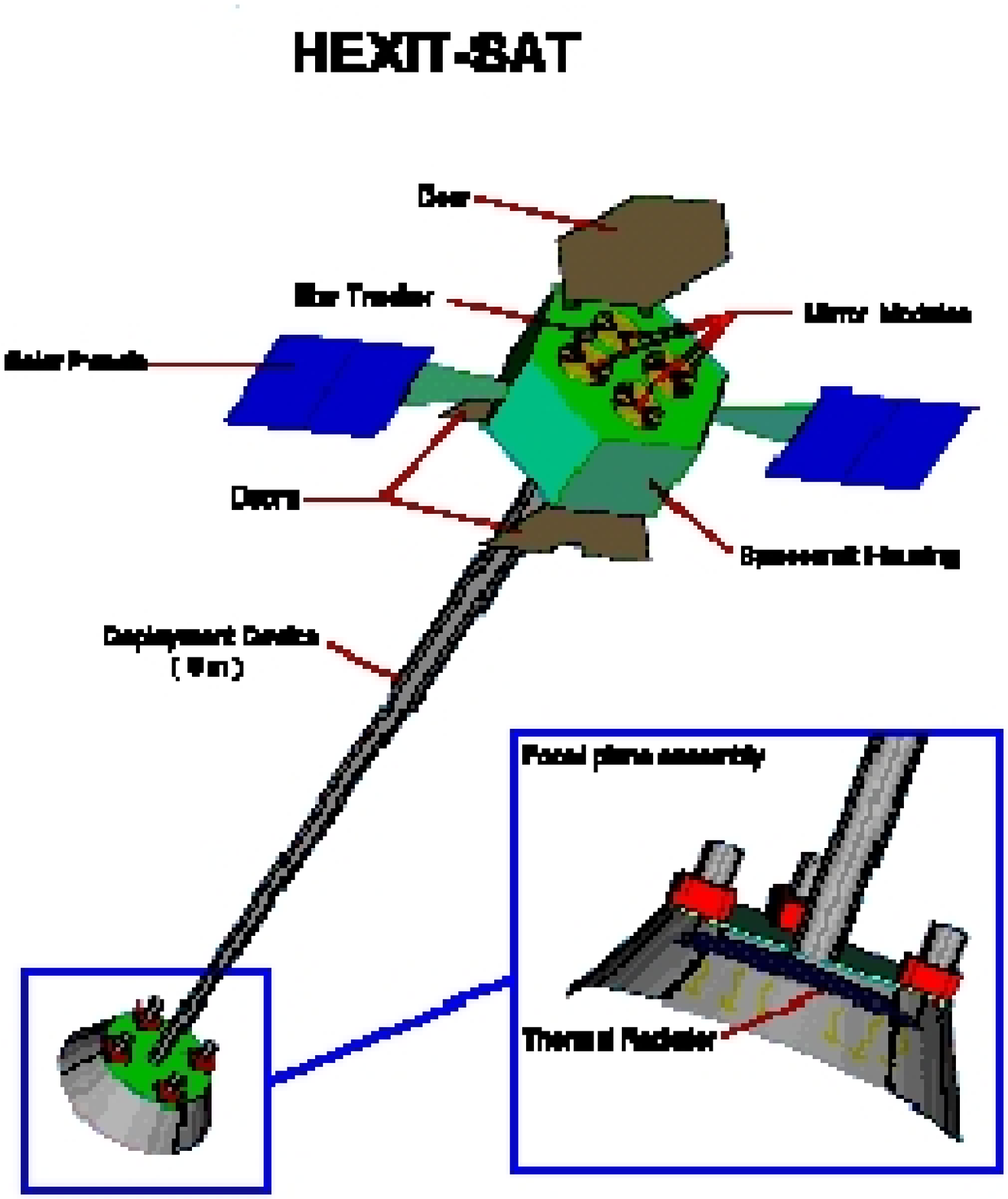}
\includegraphics[height=7cm]{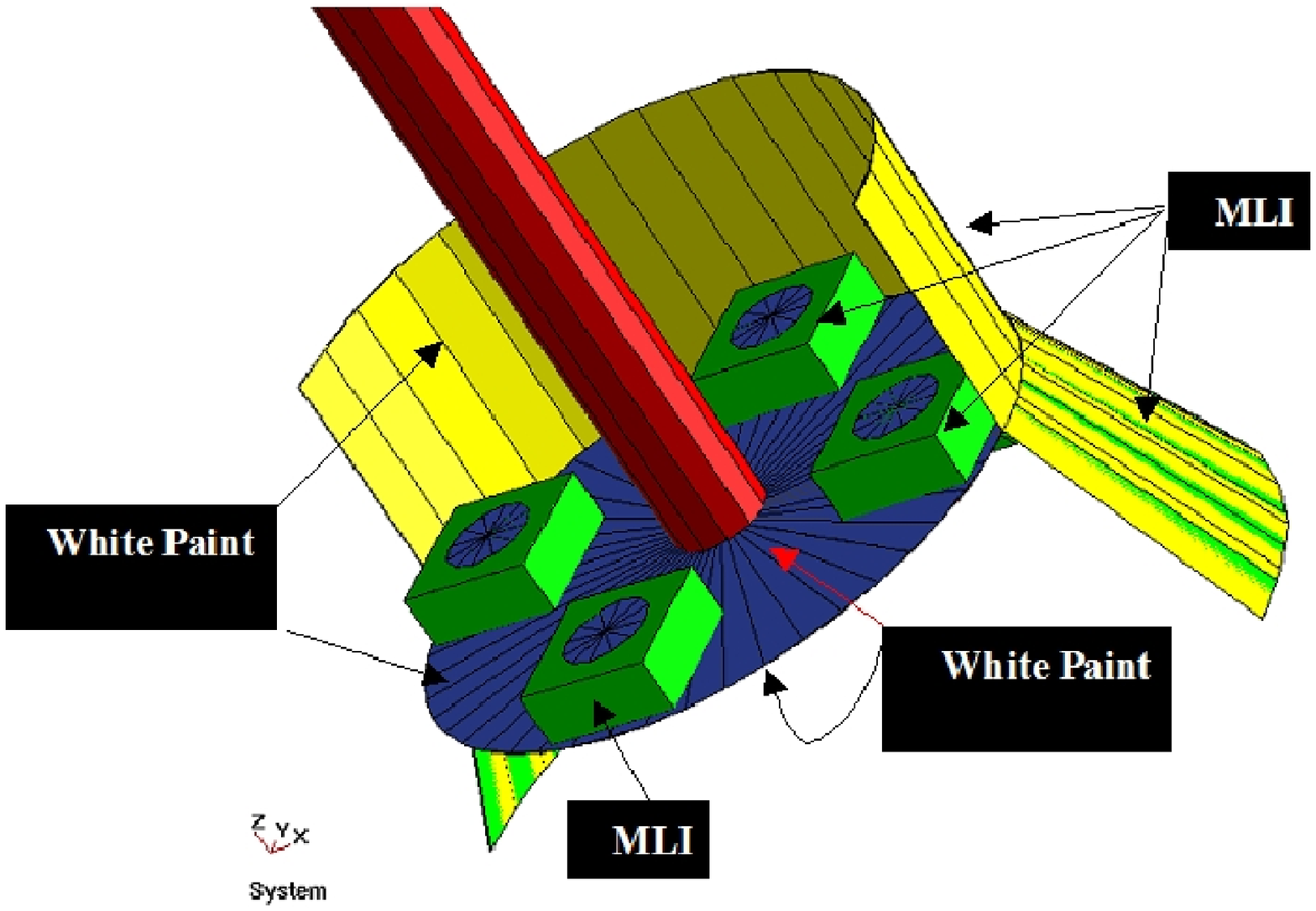}
\end{tabular}
\end{center}
\caption[example] 
{ \label{fig:sat} a), left panel: satellite Concept. The following
reference frame is adopted: the Z-axis is the long axis of the boom,
positive from focal plane assembly to mirror module assembly, the
X-axis is the rotation axis of the solar panels, the Y-axis completes
an orthonormal triad, positive in the semiplane containing the sun.
b), right panel: conceptual scheme of focal plane detectors and
coolers}
\end{figure}

Alenia Spazio performed a preliminary assessment of some aspects of
the proposed experiment's satellite implementation. A 1200 kg class
spacecraft is envisaged, whose service elements are be drawn
from the complement of ASI's standard platform PRIMA. For very low
background, as proven by BeppoSAX, a circular, equatorial orbit at
600-km mean altitude is demanded. Launch into near-equatorial orbit is
expected to be a standard provision of low-cost launch vehicles such
as Vega, Soyuz or PSLV.  Fig. \ref{fig:sat}a) shows the overall satellite
configuration concept. The satellite is made up of the focal plane
assembly (FPA) module and the spacecraft platform, with four embedded
mirror modules, separated by an 8m long deployable boom. The satellite
attitude is inertial for the duration of at least one orbit
revolution. Pointing constraints apply with respect to the sun
(30 degrees), the Earth limb (30 degrees), and the Moon (5 degrees). For
any pointing direction compatible with these constraints, a rotational
degree of freedom, around the long axis of the spacecraft (X), is
available and is used for thermal and power reasons.  Initial
investigations concerned the FPA thermal design and the boom design
concept.

\subsection{Focal Plane Assembly thermal design}

A preliminary analysis of the FPA configuration and thermal design was
performed.  The purpose of this exercise was to find the minimum
temperature achievable in the sensors in the selected orbit and with a
suitable thermal design. In the first part of the study, documented in
this paper, passive approaches were addressed. The objective was to
find a thermal design and configuration providing the lowest possible
temperatures for the sensors, without the use of thermoelectric
(Peltier) coolers. Once a configuration with suitably low base
temperature is found, the addition of Peltiers will be studied.

A number of thermal design solutions were addressed. The configuration
eventually selected is shown in Fig. \ref{fig:sat}b.  The four detector
assemblies are placed flush to a circular radiator plate, white
painted on both sides. Sun shields protect the radiator from direct
Sun illumination (half-cylindrical shield on the upper side,
half-conical shield on the lower side). The shields are covered with
multi-layer insulation (MLI) on the Sun side, and white-painted on the
internal side. 

\begin{figure*}
\begin{center}
\begin{tabular}{c}
\includegraphics[height=7cm]{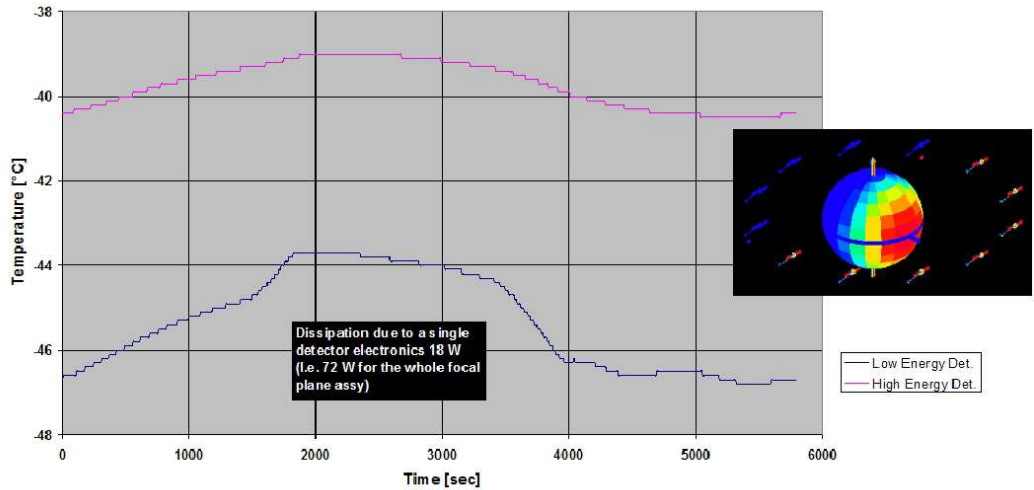}
\end{tabular}
\end{center}
\caption[example] 
{ \label{fig:dissipation}  Detector temperature vs. Equinox, 
satellite pointing 45degrees from North Pole. Full detector 
dissipation}
\end{figure*}

The FPA temperature is affected by the Sun and Earth (albedo and IR)
heat inputs, hence very sensitive to the spacecraft attitude, its
orientation with respect to the Sun and the Earth. A number of
different attitudes and configurations were simulated, and the
selected configuration of Fig. \ref{fig:sat}b is the one that was found
least affected by the environment. The coldest attitude of the S/C
detector plane occurs when the satellite points at the North pole, and
the Sun is in the equator plane, hence the cross section of the
detector plane to the incident sunlight is very small. If no
dissipation is applied, the sensor temperature is determined by
Earthshine only, around -70C.  This is the purely radiative
equilibrium temperature for that configuration and represents the
minimum obtainable temperature.  Fig. \ref{fig:dissipation} shows the
results for a realistic case (full dissipation of 9W+9W applied to
each detector, equinox, satellite pointing at 45 degrees from the
equator plane), in which the effect of the Earth is considerable. The
temperature of the high-energy detector (the most exposed to the
environment) oscillates around -40C with small variation (+/-1C) at
orbit frequency. We conclude that detector temperatures significantly
lower than -20C (as required by one type of the low energy X-ray detector
under study), at aspects to the Earth +/-30 degrees, are easily
achieved by a passive thermal control approach such as that presented
above.

To provide a further drop to -90C, as required by the other candidate
for the low-energy detector, a thermoelectric cooler is envisaged. For
the cooler to work efficiently, the dissipation applied to the sensor
element must be small. A cooler applied to the detector casing is not
efficient (the environmental + dissipated head load is too high). The
solution involves finding a good thermal contact of the cold side of
the cooler with the sensor, while the hot side has a direct thermal
link (independent of that of the casing) to the radiator. The thermal
contact must occur on the side of the sensor, to avoid impairing
the transparency of the upper sensor to hard X--rays. For a typical
Peltier, a sensor heat load of 1W and a design temperature drop of
around 50 C, the power consumption of the Peltier would be less than
20W, which could be handled by the radiator. Principle solutions have
been identified, and development of a prototype FPA has been proposed
to test the design.

\subsection{Deployable boom design}
 
\begin{figure}
\begin{center}
\begin{tabular}{cc}
\includegraphics[height=6.cm]{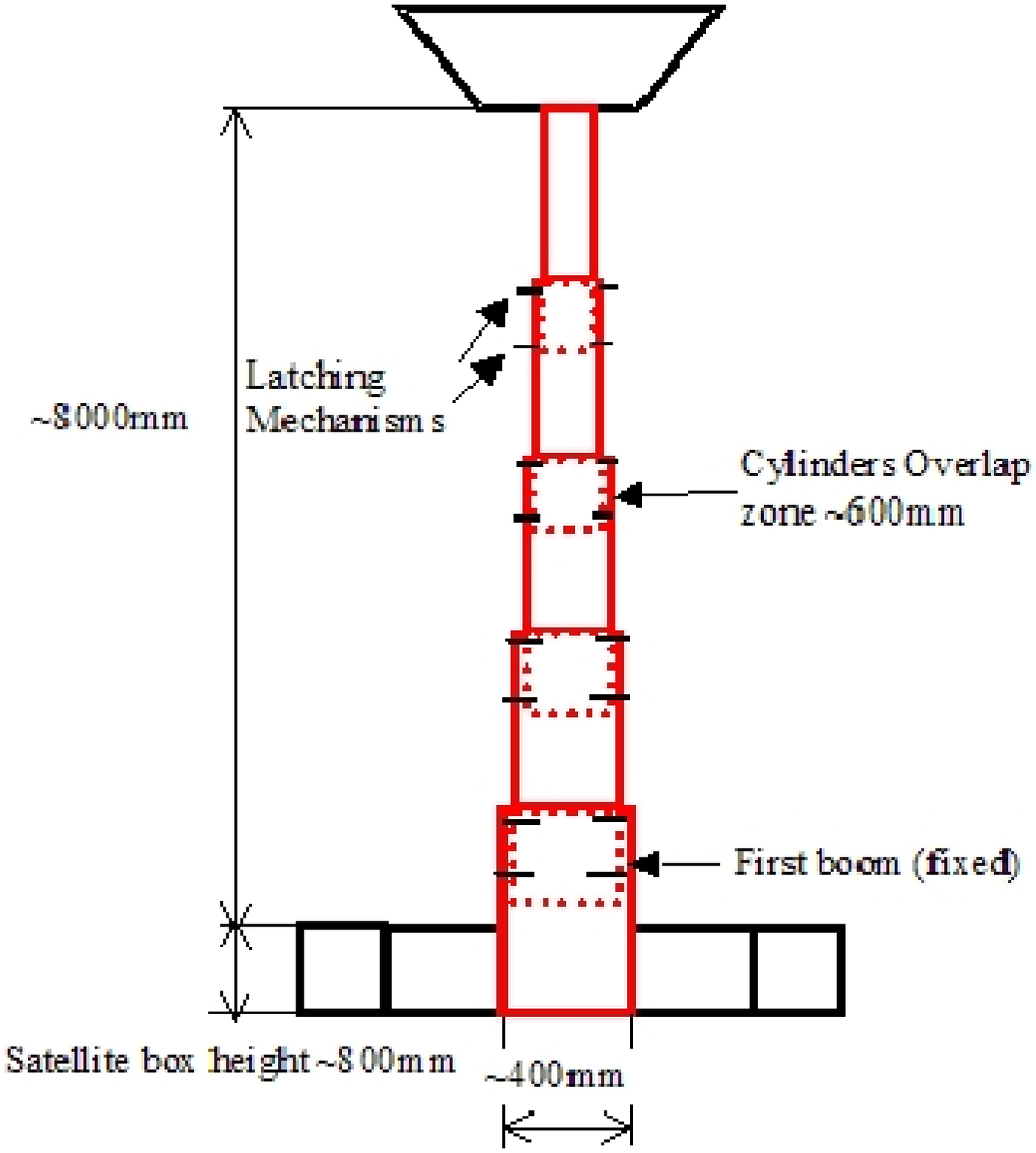}
\includegraphics[height=6.cm]{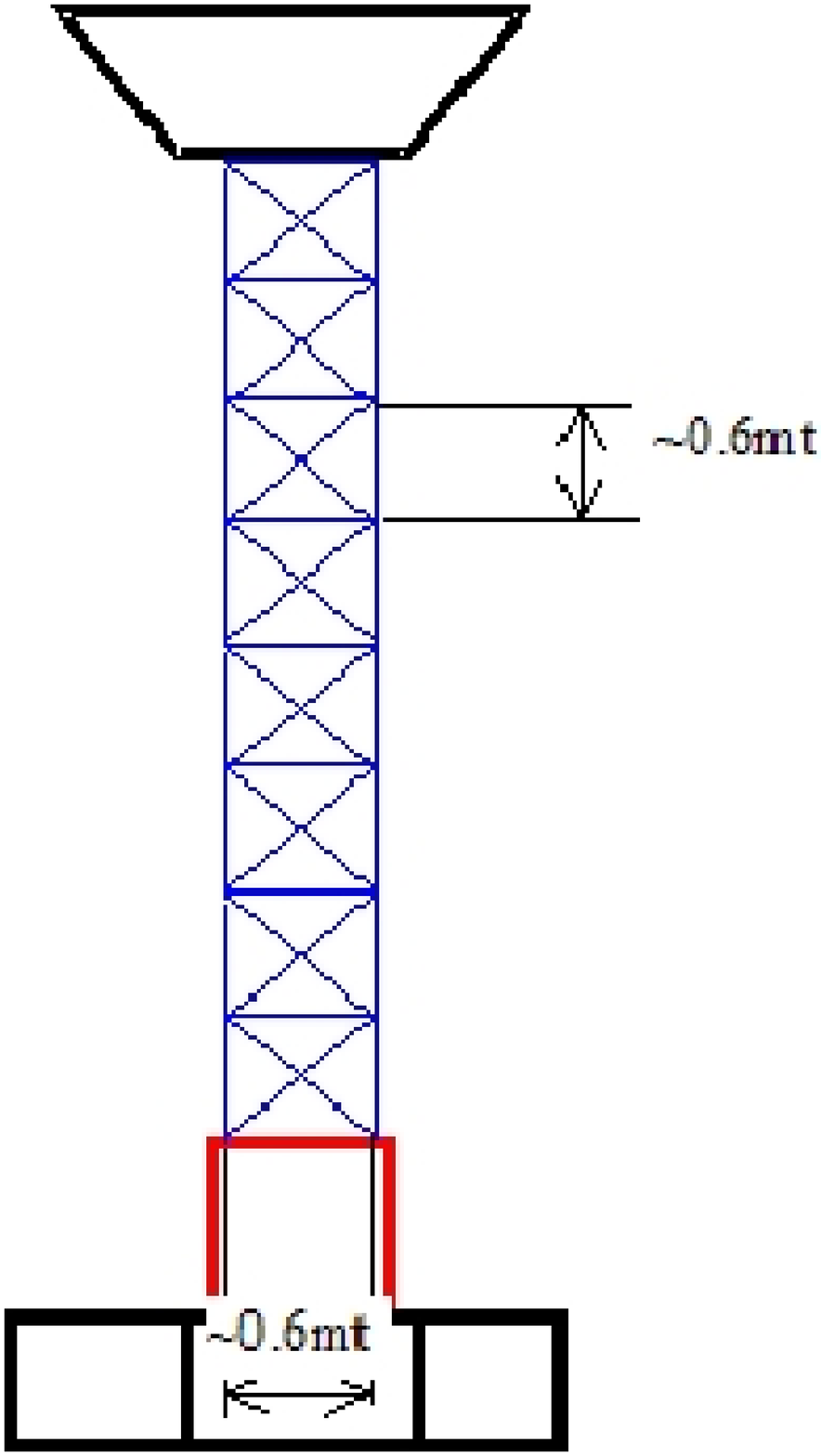}
\end{tabular}
\end{center}
\caption{
\label{fig:boom} Two possible design for the extensible boom/truss.}
\end{figure}

The boom must offer a reliable way of separating the FPA from the
mirror modules, from an initial stowed distance of about 2 m to the
operating focal length of 8 m. The driving performance requirements
concern the boom dimensional stability over the time scale of an
observation, under large temperature gradients (both axial and
circumferential). The focal length must not change by more than 2 mm
and the alignment between the optics reference axis and the focal
plane reference axis must be stable within 1 arcmin. 
The relative position of the FPA with respect to the optics will be
monitored by means of an optical alignment system. For each detected
photon a positional correcting factor will be determined. Design
requirements include small stowed volume, accommodation of cable
harness, stiffness under launch loads by a suitable restraint system,
stiffness in flight under loads from repointing and orbit maintenance
maneuvers, and more. A number of design concepts were assessed,
ranging from inflatable structures to telescopic booms, to coilable
masts to truss-like structures. As a result of the initial
investigation, two designs were selected for further study: a
lightweight telescopic boom made up by seven nested cylinders, and an
articulated truss structure made up of 12 bays (Fig. \ref{fig:boom}).
Both structures would be made of carbon fiber for high stiffness,
thermal conductivity and thermal stability. Both structure types have
heritage in space missions, although the application at hand has more
demanding requirements than most (loads, stability). A
fully-fledged design exercise is as yet outstanding, but no
feasibility problems are expected.

\acknowledgments     
The original matter presented in this paper is the result of the effort of
a large number of people from about ten  INAF Observatories and Institutes
and Italian Universities and from Alenia Spazio, which we all warmly
thank.



\end{document}